\documentclass[aps,prb,twocolumn,showpacs,superscriptaddress,groupedaddress,floatfix]{revtex4-1}
\usepackage[english]{babel}
\usepackage{amsmath,graphicx,amssymb} 
\usepackage{times}
\usepackage{epsfig} 
\usepackage{epstopdf} 
\usepackage{color}
\usepackage{color} 

\usepackage{hyperref}
\hypersetup{
colorlinks=true,final=true,
        linkcolor=red,
        citecolor=blue,
        filecolor=blue,
        urlcolor=blue,
}
\begin{document} 
\title{Quantum critical local spin dynamics 
near the \\ 
Mott metal-insulator transition in infinite dimensions}

\author{Nagamalleswararao Dasari$^{1}$}\email{nagamalleswararao.d@gmail.com} 
\author{N. S. Vidhyadhiraja$^{1}$}
\author{Mark Jarrell$^{2,3}$} 
\author{Ross H. McKenzie$^{4}$}\email{r.mckenzie@uq.edu.au}\homepage{URL: condensedconcepts.blogspot.com}
\affiliation{$^{1}$Theoretical Sciences Unit, Jawaharlal Nehru Centre For
Advanced Scientific Research, Jakkur, Bangalore 560064, India.}
\affiliation{$^{2}$Department of Physics $\&$ Astronomy, Louisiana State University, Baton Rouge, LA 70803, USA.}
\affiliation{$^{3}$Center for Computation $\&$  Technology, Louisiana State University, Baton Rouge, Louisiana 70803, USA}
\affiliation{$^{4}$School of Mathematics and Physics, University of Queensland, Brisbane 4072, Australia.}

                   
\begin{abstract}
Finding microscopic models for metallic states that exhibit quantum critical properties is a major 
theoretical challenge.  We calculate the dynamical local spin susceptibility $\chi(T,\omega)$ for 
a Hubbard model at half filling using Dynamical Mean-Field Theory, which is exact in infinite 
dimensions.  Qualitatively distinct behavior is found in the different regions of the phase diagram: 
Mott insulator, Fermi liquid metal, bad metal, and a quantum critical region above the finite 
temperature critical point.  The signature of the latter is $\omega/T$ scaling where $\omega$ is 
the frequency and $T$ is the temperature. Our results are consistent with previous results showing 
scaling of the dc electrical conductivity and are relevant to experiments on organic charge transfer 
salts.
\end{abstract}

\maketitle 

\section{Introduction}

A wide range of materials exhibit properties characteristic of strongly correlated electrons. 
Materials include transition metal oxides \cite{Imada:1998}, cuprates \cite{Keimer:2015}, 
iron-based superconductors \cite{Analytis:2014}, heavy fermion compounds \cite{Gegenwart:2008}, 
and organic charge transfer salts \cite{Powell:2011}.  They exhibit emergent quantum states of 
matter such as unconventional superconductors, spin liquids, and non-Fermi liquid metals.
A major challenge is to understand these metallic states which have properties quite distinct 
from those of simple elemental metals that can be described by Landau Fermi liquid theory. 
These unusual metallic states occur in close proximity to a Mott insulating phase \cite{Imada:1998}
and/or to a quantum critical point\cite{Gegenwart:2008}.  The concept of quantum criticality may 
be a useful organizing principle\cite{Sachdev:1999,Sachdev:2000,Sachdev:2011}.

The Hubbard model is one of the mostly widely studied effective Hamiltonians for strongly 
correlated electron systems. At the level of Dynamical Mean-Field Theory (DMFT) 
\cite{w_metzner_89a,v_janis_91,a_georges_92a,m_jarrell_92a,t_pruschke_95,Georges:1996}, 
at half filling and zero temperature there is a first-order phase transition between metallic and 
Mott insulating phases as the interaction strength $U$ is increased.  Near half filling, 
using the non-crossing approximation and quantum Monte Carlo, Pruschke et al.\ identified a region of 
anomalous transport\cite{m_jarrell_94,t_pruschke_95}.  It is characterized by linear-in-temperature 
resistivity, which corresponds to $\omega/T$ scaling, at high temperatures, crossing over to a 
Fermi liquid at lower $T$.  The crossover scale between these regimes vanishes as half filling is 
approached, and the slope of the linear in $T$ resistivity varies like $1/x$, where $x$ is the 
doping.   More recently,  Dobrosavljevic et al. identified a broad region of the $T-U$ phase 
diagram displaying $\omega/T$ scaling in the half filled model.  They associated this behavior with a 
quantum critical point\cite{Terletska,Vucicevic:2013,Vuvcivcevic:2015}.  Furthermore, similar scaling 
was found in experimental data for three different organic charge transfer salts that exhibit a 
critical point for the Mott transition in the temperature-pressure phase diagram \cite{Furukawa:2015}. 
In this paper we show that the local spin dynamics of the Hubbard model calculated with DMFT 
exhibits $\omega/T$ scaling that is characteristic of quantum criticality.

{\it Quantum criticality and $\omega/T$ scaling.}  Varma et al.\ \cite{Varma:1989,Varma:2016} 
showed that many of the anomalous properties of the metallic phase of cuprate superconductors at 
optimal doping can be described as a  marginal Fermi liquid with a spin fluctuation spectrum 
that exhibits $\omega/T$ scaling.  Finding concrete realistic theoretical microscopic fermion 
models that exhibit such scaling has proven challenging.  Simulations of the two-dimensional 
Hubbard model reveal a quantum critical point at finite doping below a fan shaped region of 
marginal fermi liquid character in the self energy\cite{n_vidhyadhiraja_09,e_khatami_10}.  
There are several reviews of quantum criticality\cite{Sachdev:1999,m_voita_03,d_belitz_04,
p_coleman_05,j_lohneysen_07,q_si_10,Sachdev:2011,q_si_14}.  Sachdev has reviewed several 
spin and boson models\cite{Sachdev:1999} that exhibit $\omega/T$ scaling in the quantum critical 
region, associated with a quantum critical point.  In such systems, the temperature itself is the relevant 
low energy scale, rather than any scale in the model.  For example, for the transverse 
field Ising model in one dimension (page 73 of Ref. \onlinecite{Sachdev:1999}), the spin dephasing rate 
$\Gamma = 0.4 T$.  For the two-dimensional $O(N \geq 3)$ rotor model in the large-$N$ limit,
$\Gamma = 0.94 T/N$ (page 142 of Ref. \onlinecite{Sachdev:1999}).  Parcollet and Georges \cite{Parcollet:1999} 
considered a particular limit of a random Heisenberg model which had a spin liquid ground state 
and the dynamical spin susceptibility $\chi''(T,\omega)$ exhibited a form consistent with that 
conjectured in the marginal Fermi liquid scenario.  Neutron scattering measurements found that 
the dynamic spin susceptibility exhibits $\omega/T$ scaling for an insulating antiferromagnetic 
spin chain compound\cite{Lake:2005} and a Kagome lattice material \cite{Helton:2010}.  Quantum 
criticality has been found for a Kondo boson-fermion model \cite{Zhu:2004,Kirchner:2008}, 
motivated by $\omega/T$ scaling seen in neutron scattering experiments on several heavy fermion 
metals \cite{Schroder:2000,Gegenwart:2008}.  Specifically, inelastic neutron scattering gives 
the following $\omega/T$ scaling, for the wave-vector dependent susceptibility, 
$ \chi''(\omega, \vec{q})^{-1}= T^a F( \omega/T) + \chi'(\omega=0, \vec{q})^{-1}  $ where 
$\vec{q}$ is the wavevector and the exponent $a =0.75$.

\begin{figure}[htb] 
\centering 
\includegraphics[angle=0,width=1.0\columnwidth]{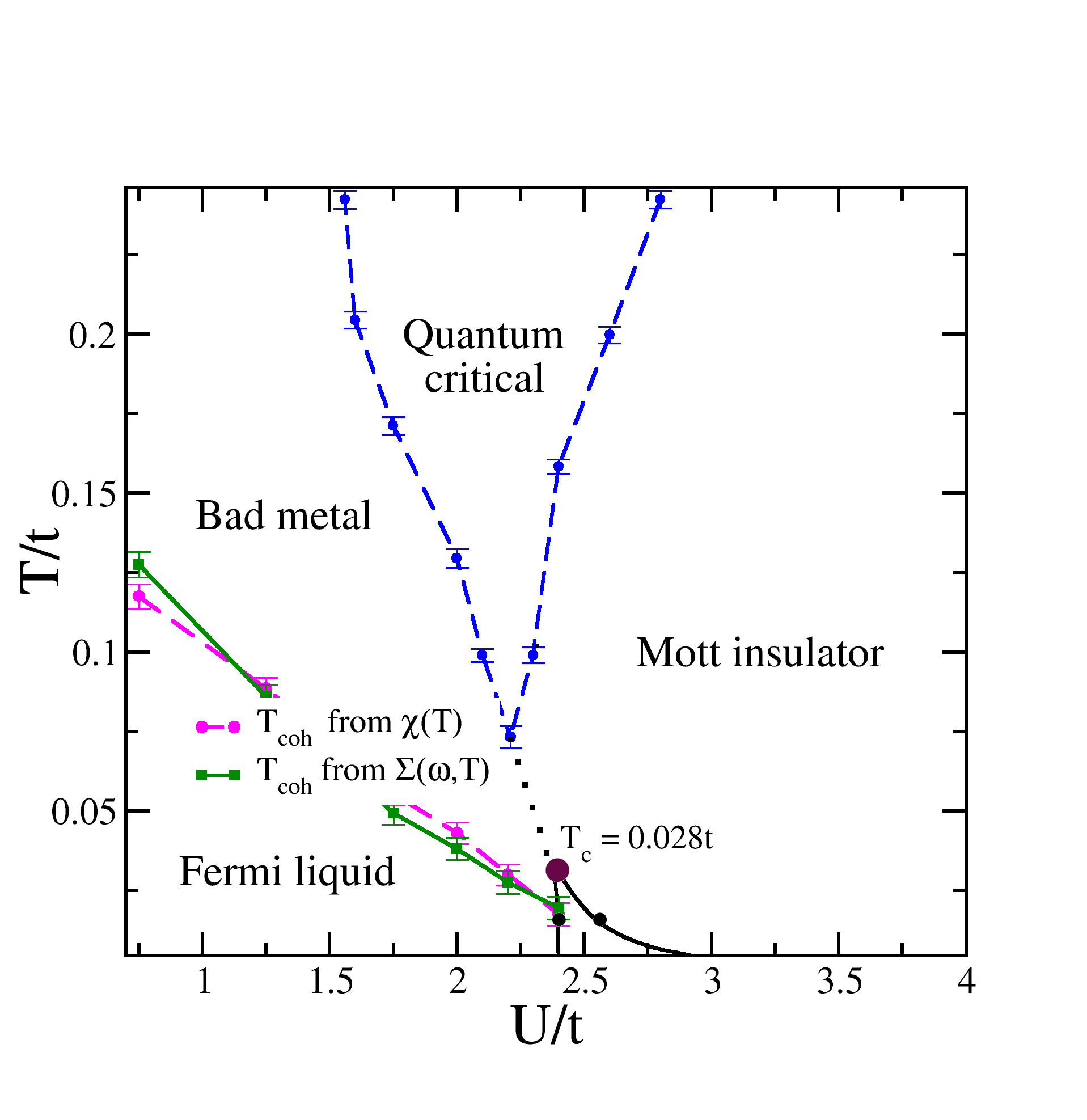}
\caption{
Phase diagram determined from the dynamical spin susceptibility.  In the quantum critical region 
$\chi''(\omega)$ exhibits $\omega/T$ scaling with a spin relaxation rate that is proportional to 
the temperature. There is a finite-temperature critical point for the Mott metal-insulator
transition ($T_c=0.028t$).  In the Mott insulating phase $\chi''(T,\omega)/\omega$ tends to a 
delta function peak as the temperature tends to zero.  In the Fermi liquid phase the spin
relaxation rate is independent of temperature.
The coherence temperature for the Fermi liquid was defined in two independent ways.  The first 
is where the static spin susceptibility becomes temperature dependent, and the second where the 
imaginary part of the one-electron self energy $\Sigma''(\omega=0,T)$ deviates from a $T^2$ 
dependence (see Figure Fig.~\ref{fig:Tcoh}  in the Appendix).  The black lines define the 
co-existence region of the metal and Mott insulator, and the critical point, as determined in
Ref.~\onlinecite{Werner:2007}.  The black circles are our results.  The blue symbols are the 
boundary of the quantum critical regime determined by the electron spin relaxation rate plotted 
in Fig.~\ref{fig:gamma}.
}
\label{fig:phased}
\end{figure}

Our results are summarized in the phase diagram shown in Figure \ref{fig:phased}.  This diagram 
is deduced from the dynamical local spin susceptibility and is similar to that previously found 
from scaling of the dc electrical conductivity near the critical point for the metal-insulator 
transition in the half-filled model\cite{Terletska,Vucicevic:2013}.  Specifically, there is a 
quantum critical regime above the critical point; the signature is that the dynamical local spin 
susceptibility exhibits $\omega/T$ scaling.  The local spin relaxation rate is linear in temperature, 
with a value $\Gamma \simeq 0.4 T$.  The occurrence of quantum critical properties in both the spin 
and charge sectors is consistent with recent work showing they are strongly coupled near the Mott 
transition \cite{Lee:2016}.

\section{Model Hamiltonian}

We study the single band Hubbard model on the Bethe lattice in infinite dimensions and at half filling.
\begin{equation}
H=-t \sum_{ \langle{i,j}\rangle,\sigma} (c_{i,\sigma}^\dagger c_{j,\sigma}+ h.c.) +
U\sum_i n_{i\uparrow}n_{i\downarrow} 
\end{equation}
It involves two parameters: $t$ the nearest neighbor hopping integral and $U$ the Coulomb repulsion 
energy for two electrons on the same lattice site.  The non-interacting ($U=0$) density of states is 
semi-circular with a full bandwidth $W=2t$. DMFT is used to calculate the properties of the 
model \cite{Georges:1996}.  In the limit of infinite dimensions or of infinite lattice connectivity
DMFT is exact.  We do not allow for symmetry breaking such as antiferromagnetism.  Previously it has 
been shown that the metallic and Mott insulating phases co-exist in the range, $U_{c1} < U < U_{c2}$, 
where $U_{c1}=2.4 t $ and  $U_{c2}= 2.9 t$ \cite{Werner:2007}.  There is a finite-temperature  
critical point at $U_{c}=2.4 t $ and $T_c=0.028 t$. Our results at $\beta \equiv 1/T = 70/t$
are consistent with this earlier work (compare Figure \ref{fig:phased}).

\section{Method}

The hybridization expansion version of the continuous time quantum Monte-Carlo (CTQMC) \cite{Gull:2011} is 
used as a DMFT impurity solver to calculate the spin dynamics at finite temperature.  The main advantages 
of this method are that it is numerically exact and the fermionic sign problem does not occur until very 
low temperatures in the Fermi liquid regime.  The vertex corrected local spin susceptibility 
$ \chi(\tau) \equiv \langle S_z(\tau) S_z(0) \rangle $ is computed at imaginary times and then Fourier 
transformed to Matsubara frequencies. 

In CTQMC simulations, we accumulate adjacent imaginary time one-and two-particle Green's 
function data into equal bins\cite{ALPS}. This may be done efficiently, but the data obtained is 
correlated in both Monte Carlo and imaginary time and hence may be problematic for analytic 
continuation via the maximum entropy method\cite{Jarrell:1996,m_jarrell_91b}. By increasing 
the bin size, we may reduce correlations between adjacent bin averages, ensuring that the binned 
data has a Gaussian distribution.  We can quantify this by fitting the histogram of the binned data 
to a Gaussian form and by calculating the third and fourth moments of the histogram to ensure that 
they are very small ($\sim$ 10$^{-1}$ to 10$^{-2}$).  However, correlations between errors of 
the Green’s function at adjacent time slices remain, and are characterized by the off-diagonal
elements of the covariance matrix (C).  To remove these correlations, we diagonalize the covariance 
matrix with a unitary transformation $U$
\begin{equation}
U^{-1} C U = {\sigma^{'}_i}^2 \delta_{ij}\\.
\end{equation}
We then rotate the data (G) and kernel (K) into this diagonal representation $K^{'}$ = $U^{-1} K$, 
$G^{'} = U^{-1} G$, where we may carry out the maximum entropy calculations on independent samples. 

Empirically, Jarrell et al.\cite{Jarrell:1996} find that accurate calculations 
of the covariance requires that the number of bins must be chosen such that $N_{bins} \geq 2L $, 
where $L$ is the number of required independent eigenvectors. In our calculations, we use
$N_{bins}=300$ and $L=15$. We also perform calculations in the critical region by increasing 
the number of bins from 300 to 1000 to ensure robustness and observe that the change in spin-relaxation rate $\Gamma$ is 
very marginal ($\sim 10^{-3}$). Hence, in this paper, we show data obtained for 300 bins.

As the default model for the analytic continuation we use the closed analytical form results of 
Salomaa \cite{Salomaa:1976} for the resonant level model (Anderson single impurity model with $U=0$). 
Given the data, in the Salomaa model the parameter for the width of the spectral density is chosen 
such that it maximizes the posterior probability of the model\cite{m_jarrell_91b}. We also 
calculate $\chi''(T,\omega)$ by using another default model provided by Bouadim et al.\cite{bouadim}.  
We find that our results are independent of the choice of default model, suggesting that the 
analytical continuation procedure is quite robust.

\section{Spin relaxation rate}
\begin{figure}[h!]
\centering
\includegraphics[angle=0,width=1.0\columnwidth]{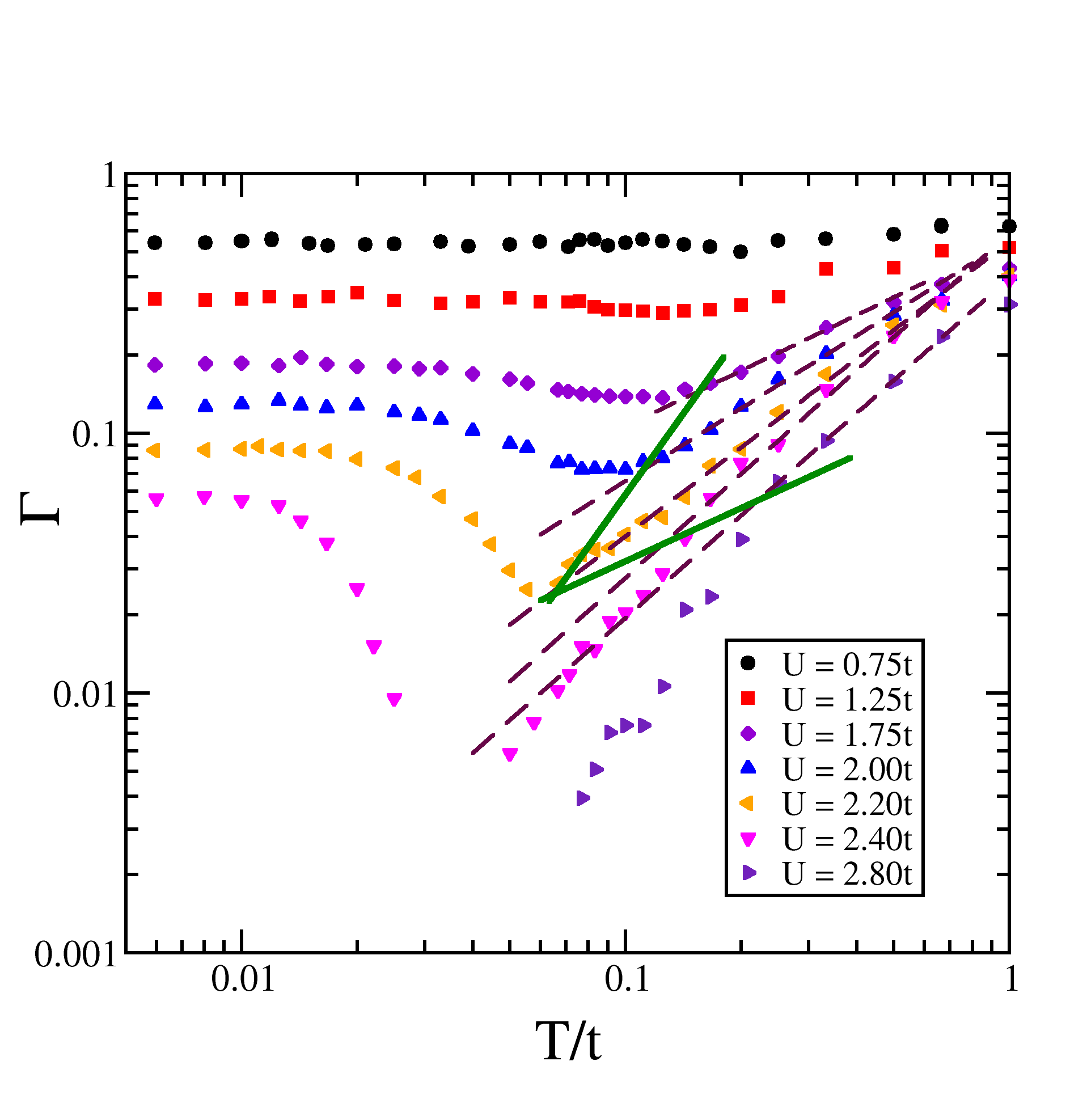}
\caption{Electron spin relaxation rate $\Gamma(U,T)$ as a function of temperature for different $U$ 
values. In the Fermi liquid regime of the metallic phase ($T < T_{coh}$) the relaxation rate is 
non-zero and independent of temperature.  This rate decreases by more than an order of magnitude as 
the Mott insulator is approached.  Above the coherence temperature $\Gamma$ decreases with increasing 
temperature, reflecting the decreasing interaction between the spins of the electrons which become 
more localized as the temperature increases.  In the Mott insulator ($U > 2.4 t$) the rate tends to 
zero as the temperature tends to zero reflecting the decoupled local moments. In the quantum critical 
regime the rate is a power law as a function of temperature. For $U=2.2t$ the rate  is approximately 
linear in temperature, $\Gamma \simeq 0.4 T$. The green lines define the boundary of the quantum 
critical region in Figure \ref{fig:phased}.
} 
\label{fig:gamma}
\end{figure}

From the dynamical local spin susceptibility, $\chi(T,\omega)\equiv \chi'(T,\omega) + i \chi''(T,\omega)$,
a spin relaxation rate can be defined by,
\begin{equation}
\Gamma(U,T) \equiv \lim_{\omega \to 0}
{\omega \chi'(T,\omega=0) \over \chi''(T,\omega)}.
\label{eq:eq1}
\end{equation}
This is similar to the (dephasing) relaxation rate defined by Sachdev for a spin model at the ordering 
wavevector (Ref. \onlinecite{Sachdev:1999}, page 73).  If $\chi(T,\omega)$ has  a simple Lorentzian or 
Drude form then $\Gamma$ corresponds to the width of the peak at zero frequency in the spectral density, 
$\chi''(T,\omega)/\omega$.  Our results for the temperature dependence of $\Gamma$ are shown
in Figure \ref{fig:gamma}.

On the metallic side of the Mott transition the signature of the crossover from a Fermi liquid to a 
bad metal (with increasing temperature above $T_{coh}$) is that $\Gamma(T,U)$ decreases smoothly from 
a small $T$ independent value below $T_{coh}$ to a temperature dependent value.  
Above this temperature the spin dynamics is weakly damped, similar 
to the localized weakly interacting magnetic moments present in the Mott insulating phase.  The latter 
was conjectured to be the character of the bad metallic state, based on the large entropy and static 
spin susceptibility found from finite temperature Lanczos calculations for the Hubbard model on the 
triangular lattice at half filling \cite{Kokalj:2013}.

Recent DMFT calculations of  charge transport properties of a doped Hubbard model \cite{Resilient} 
identified the existence of well-defined quasiparticle-like excitations [ resilient quasi-particles 
(RQPs)] well above the coherence temperature ($T_{coh}$) and their gradual extinction with the 
cross-over to the bad metallic regime ($T_{MIR}$) where the resistivity becomes comparable to the 
Mott-Ioffe-Regel limit. 
Our results suggest that the spin relaxation rate in the RQP regime behaves quite differently in 
comparison with the low temperature Fermi liquid and high temperature bad metallic regime. In 
fact, this might be relevant to the recently observed slowdown of the relaxation dynamics near 
the Mott transition in the quench dynamics of the Hubbard model\cite{quench}. Our results suggest 
the need to investigate spin dynamics in the doped case.


\section{Quantum critical scaling}

\begin{figure}[htb] 
\centering 
\includegraphics[angle=0,width=1.0\columnwidth]{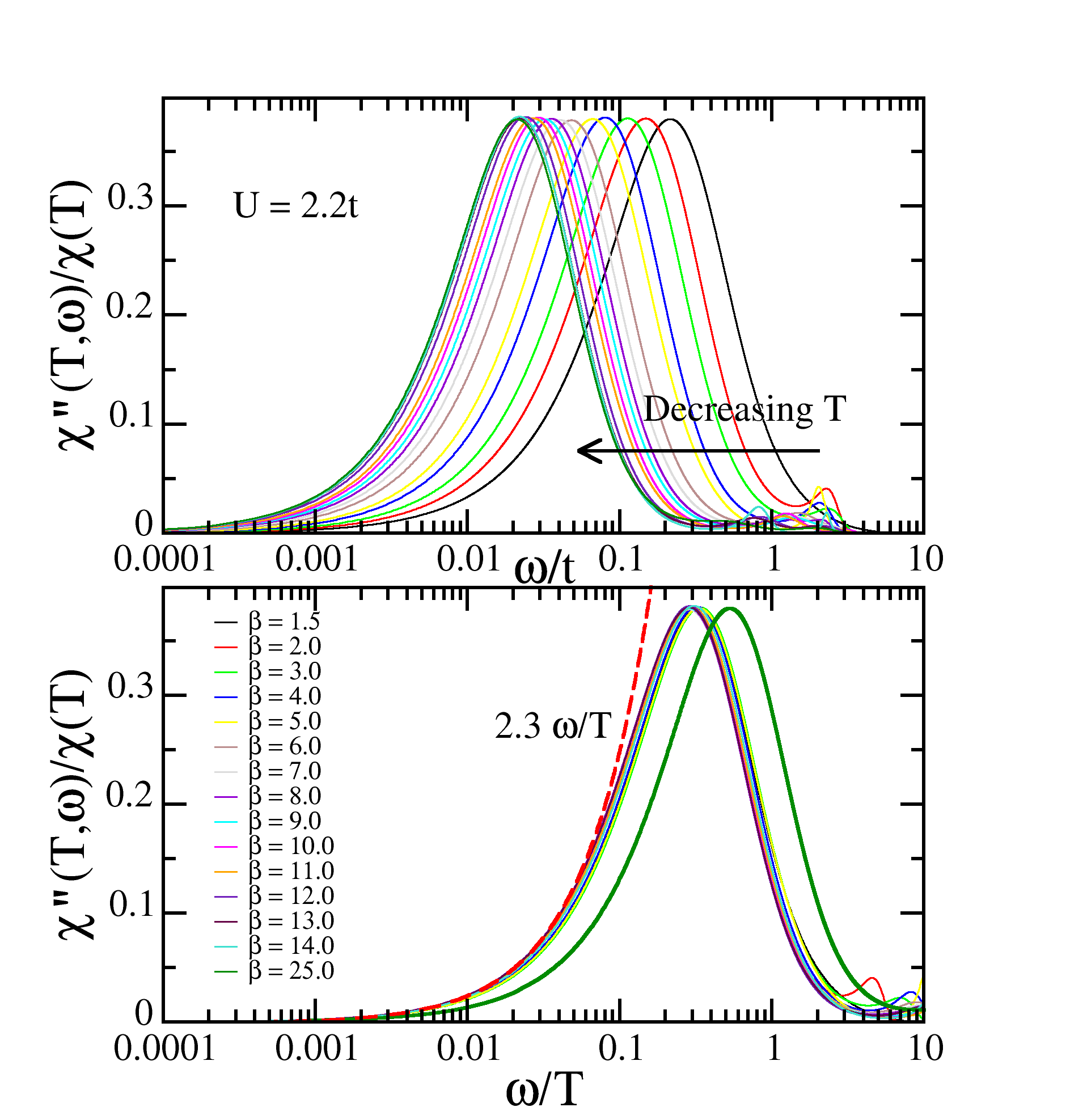}
\caption{
Quantum critical scaling of the dynamical spin susceptibility near the Mott transition. The upper 
panel shows the frequency dependence of the imaginary part of the susceptibility for $U=2.2t$ for 
a range of temperatures ($T=1/\beta$).  The lower panel shows the same data  with the frequency 
scaled by temperature. Scaling occurs for all $T \gtrsim 0.069t$, but fails for $T=0.04t$ (green 
curve) consistent with the extent of the QC region in Figure \ref{fig:phased}. The dashed line 
$2.3 \omega/T$ is a best fit to the low frequency data.
}
\label{fig:qcrit}
\end{figure}

\begin{figure}[htb]
\centering
\includegraphics[angle=0,width=1.0\columnwidth]{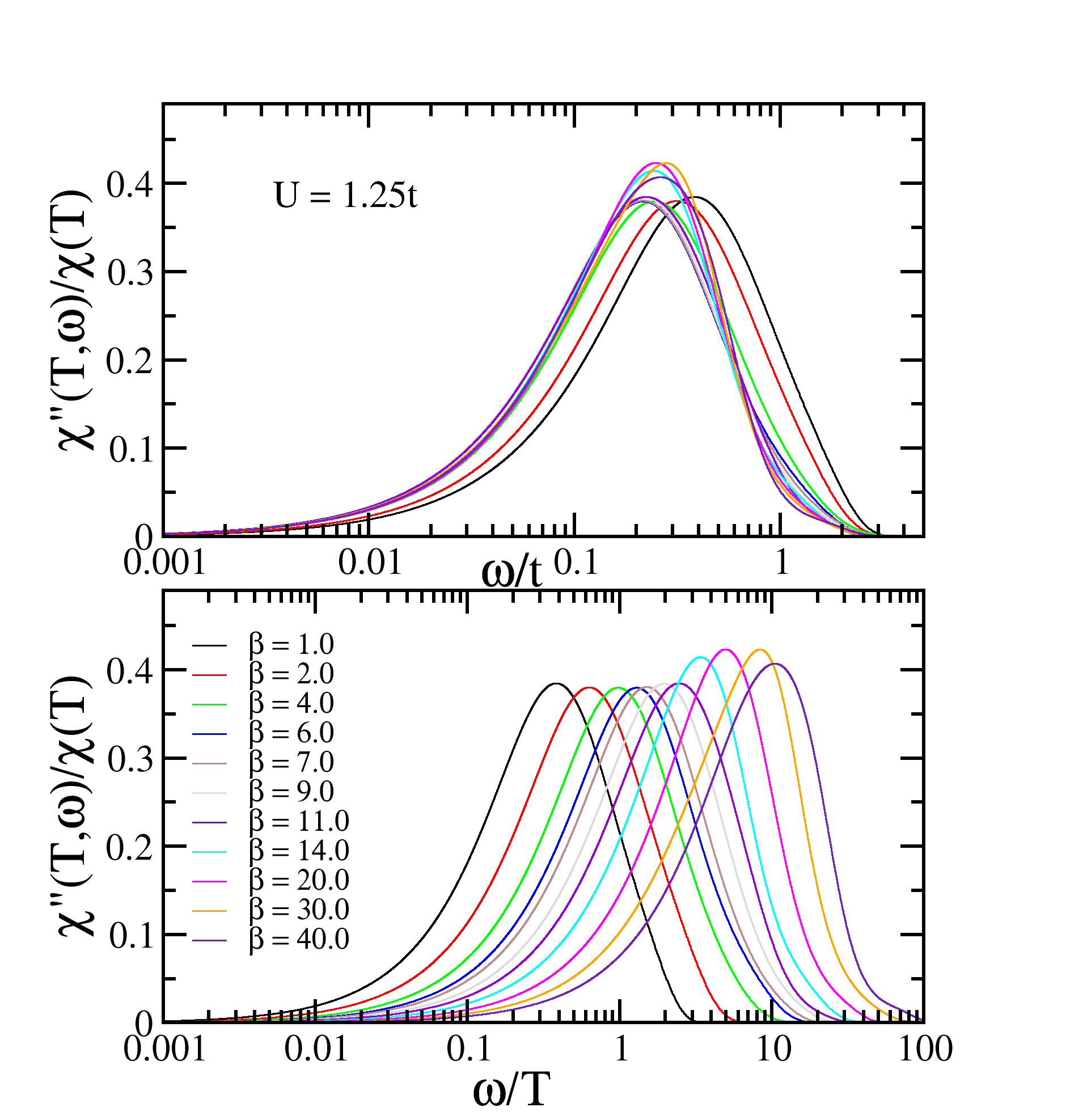}
\caption{
Frequency dependence of the dynamical spin susceptibility in the metallic phase.  The upper panel 
shows the frequency dependence of the imaginary part of the susceptibility for $U/t=1.25$ for a 
range of temperatures and on a linear scale.  The lower panel shows the same data on a log-log 
plot with the frequency scaled by temperature.  Unlike in the quantum critical regime,  
$\omega/T$ scaling is not observed.
}
\label{fig:metal}
\end{figure}

\begin{figure}[htb]
\centering
\includegraphics[angle=0,width=1.0\columnwidth]{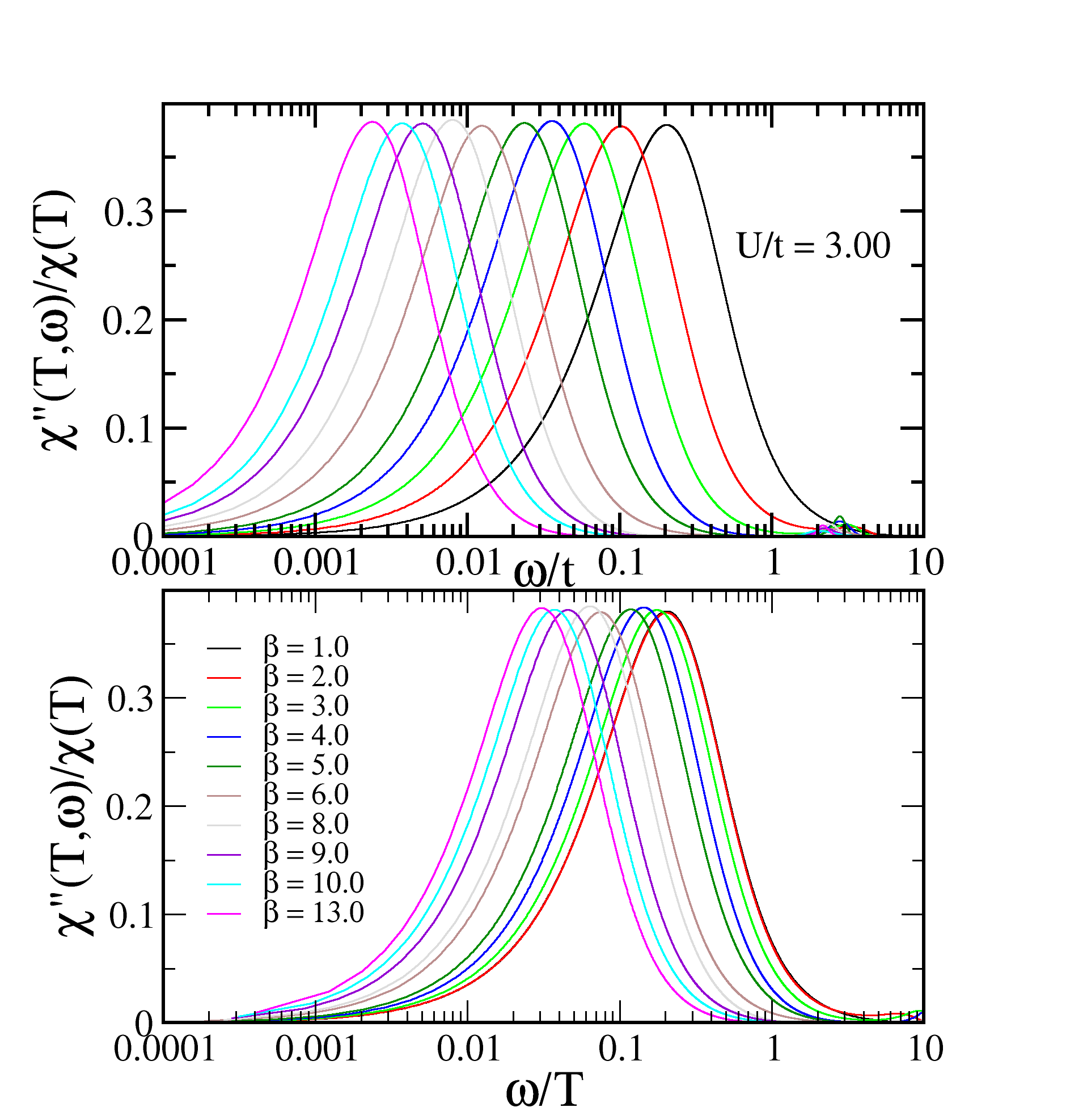}
\caption{
Frequency dependence of the dynamical spin susceptibility in the Mott insulating phase.
The upper panel shows the frequency dependence of the imaginary
part of the susceptibility for $U/t=3.0$ for a range of temperatures and on a linear scale.
Note how as the temperature tends to zero the peak width tends to zero and
that one sees features around $\omega=U$, associated with the Hubbard bands.
The lower panel shows the same data on a log-log plot with the frequency scaled by temperature.
Unlike in the quantum critical regime, $\omega/T$ scaling is not observed.
}
\label{fig:insulator}
\end{figure}

Figure \ref{fig:qcrit} shows that above the critical point $ \chi''(T,\omega)$ exhibits
$\omega/T$ scaling characteristic of quantum criticality, i.e., 
$ \chi''(T,\omega)=\chi'(T,\omega=0)F(\omega/T)$. For low frequencies the scaling function is best 
fit to a power law, $F(x)=2.3 x$.  For $T \gtrsim 0.069t$ the scaling covers about three decades 
in the ratio $\omega/T$.  For comparison, in the metallic (Figure \ref{fig:metal}) and Mott 
insulating (Figure \ref{fig:insulator}) regions such scaling clearly does not occur.  In the Mott 
insulating phase $ \chi''(T,\omega)/\omega$ tends towards a delta function peak, i.e., 
$\Gamma(T) \to 0$ as $T \to 0$ (compare Figure \ref{fig:insulator}).  This reflects the decoupled 
local moments in the Mott phase. In DMFT there is no Heisenberg antiferromagnetic exchange 
interaction between localized spins on neighboring lattice sites.
At zero temperature, the delta function peak is also clearly seen in dynamic DMRG 
calculations \cite{Raas:2009}.

\subsection{Boundary conformal field theory (CFT) scaling}

\begin{figure}[htb]
\centering
\includegraphics[angle=0,width=0.9\columnwidth]{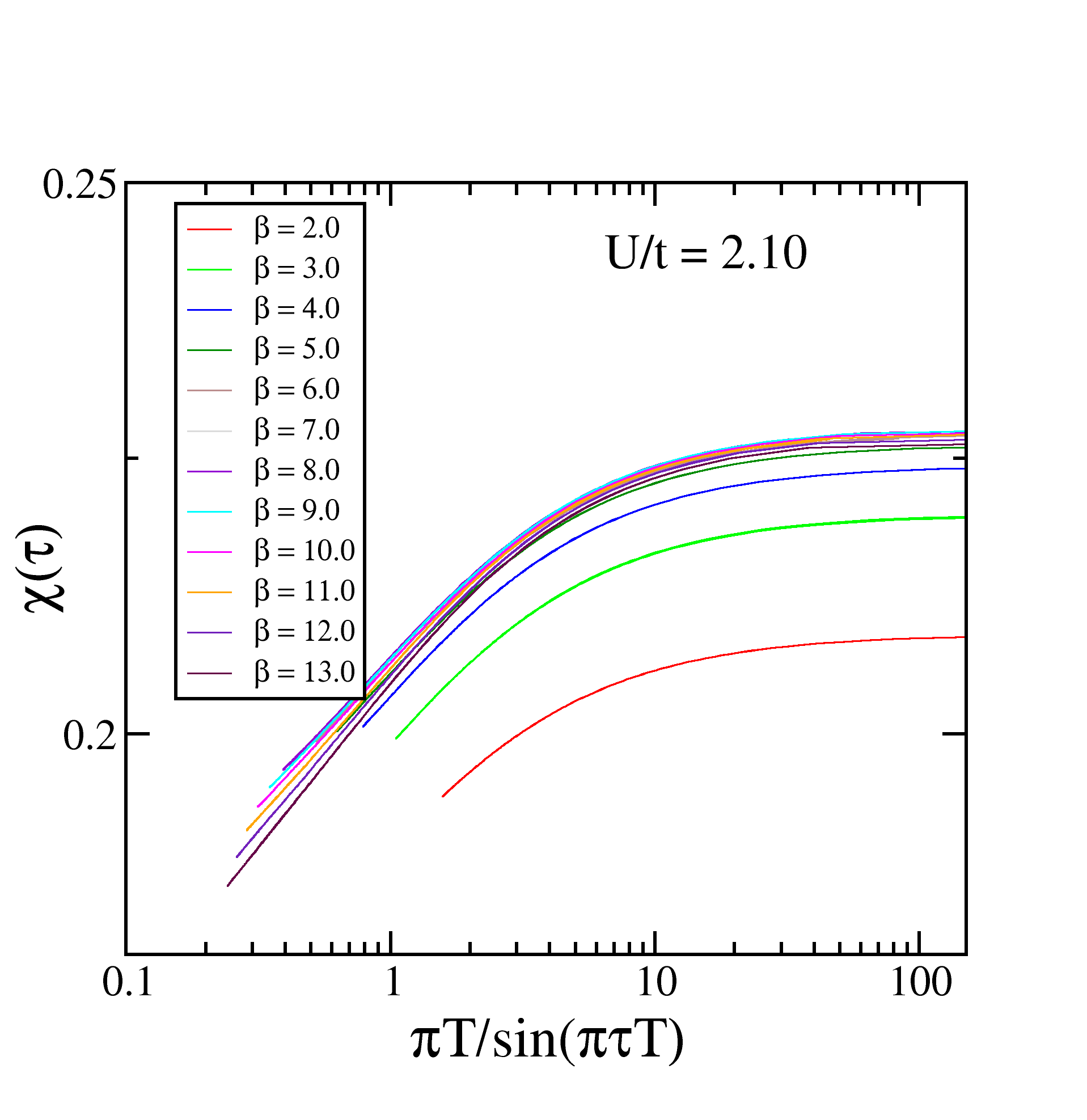}
\includegraphics[angle=0,width=0.9\columnwidth]{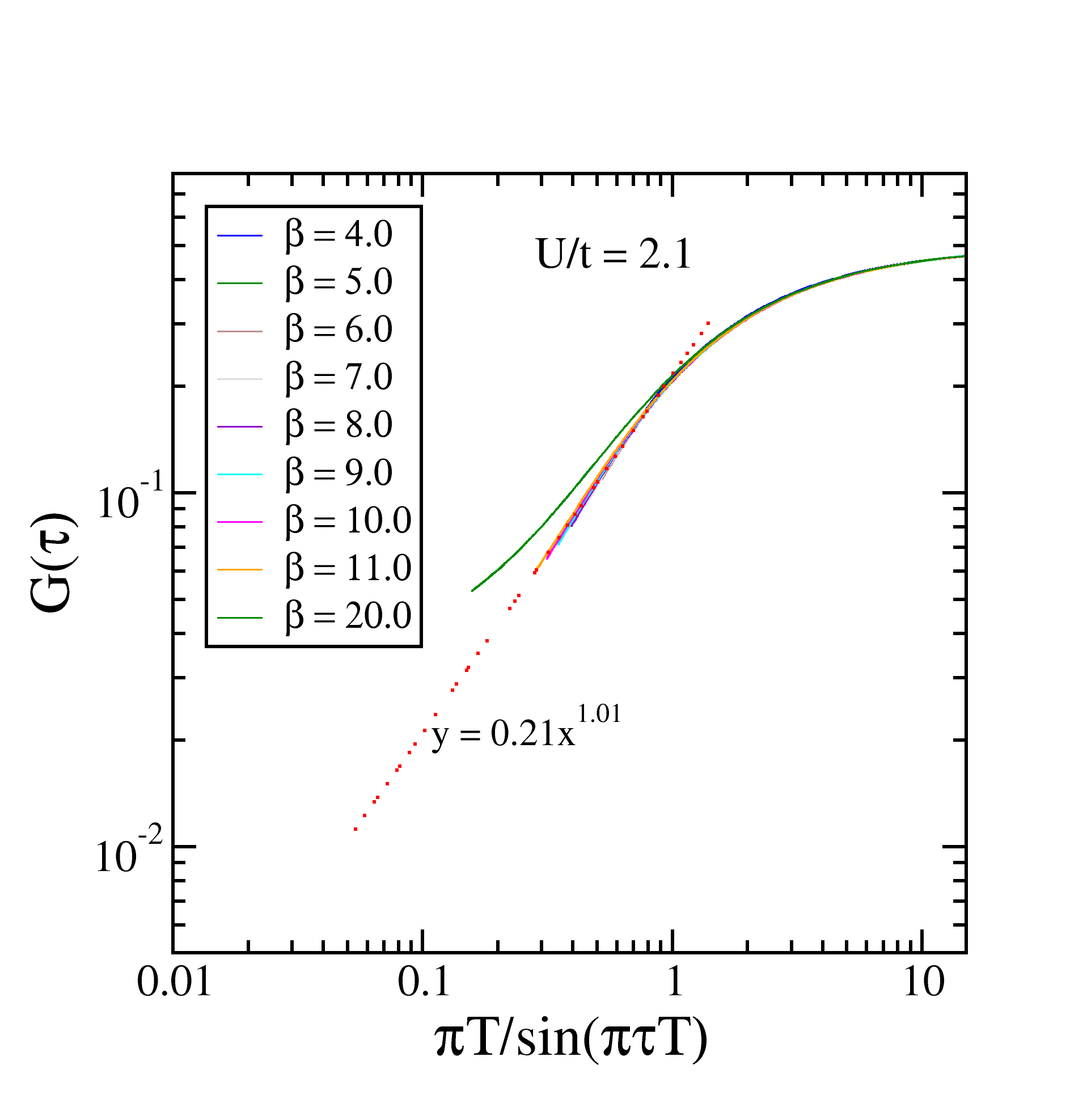}
\caption{
The upper panel shows the failure of boundary conformal field theory scaling of the imaginary time dependence 
the susceptibility $\chi(\tau)$. The lower panel shows the boundary conformal field theory scaling of the
imaginary time dependence of the one electron Green's function, $G(\tau)$. Curves are shown for $U =2.10 t$ 
and for a range of temperatures ($\beta=1/T$).  We  tried to find a temperature dependent rescaling parameter 
for the  vertical axis such that all curves of this spin susceptibility collapse on top of each other. However, 
we did not find any such rescaling parameter.}      

\label{fig:fail}
\end{figure}

Scaling with $\omega/T$  is associated with the following scaling of the imaginary-time susceptibility 
\begin{equation}
\chi(\tau) \sim
(\pi T/\sin(\pi \tau T)^{2 \lambda}
\end{equation}
where $\tau$ is the imaginary time \cite{Affleck}.  It has been found that such scaling does hold for 
a Kondo boson-fermion model \cite{Zhu:2004,Kirchner:2008,Glossop:2011}, a fractionalised Fermi liquid 
in a holographic metal \cite{Sachdev:2010}, a pseudogap Anderson model \cite{Pixley:2013}, and the 
gapped single impurity Anderson model \cite{Dasari:2015}.  We do not observe such scaling in 
$\chi(\tau)$ but do observe such scaling in the one electron local Green's function $G(\tau)$ 
(compare  Figure \ref{fig:fail}).  Our results illustrate that $\omega/T$ scaling does not necessarily 
imply the scaling characteristic of boundary CFT.

\section{NMR properties}

The most direct experimental probe of the low frequency behavior of  the  dynamical local spin susceptibility 
$\chi''(\omega)$ is through Nuclear Magnetic Resonance (NMR). In contrast, neutron scattering measures the 
dynamical susceptibility at finite wave vector.

\begin{figure}[htb] 
\centering 
\includegraphics[angle=0,width=1.0\columnwidth]{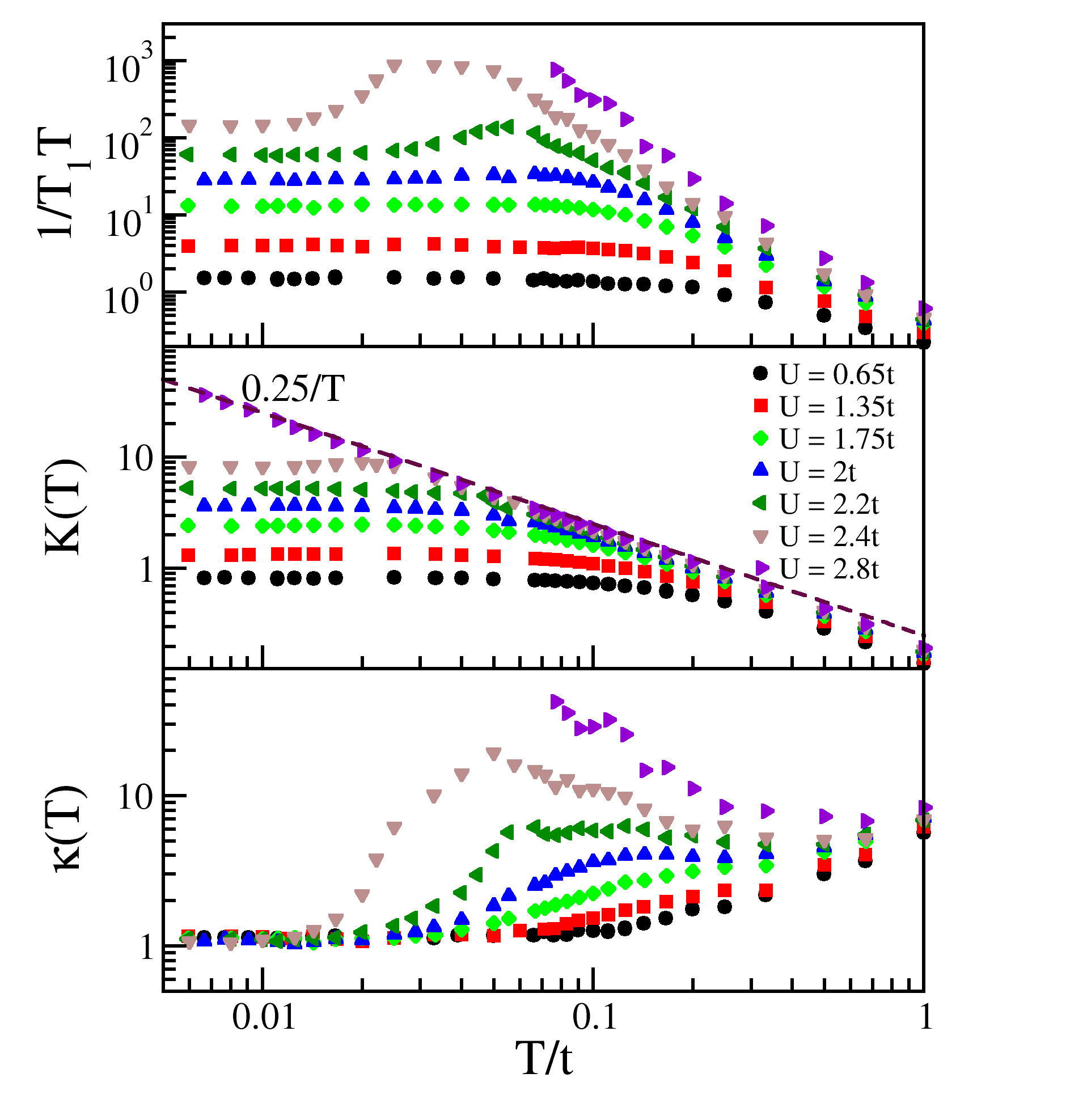}
\caption{
Temperature dependence of
NMR properties for a range of $U$ values.
The top panel  shows the nuclear spin relaxation rate, $1/T_1 T$. 
The middle  panel shows the local static susceptibility (Knight shift).
Note that both quantities are significantly enhanced as the Mott transition is approached 
on the metallic site. Below the coherence temperature $T_{coh}$, both are 
independent of temperature, characteristic of a Fermi liquid.
Well above $T_{coh}$ the static susceptibility approaches the Curie form
($\chi =1/(4T)$, shown by the purple dashed line), suggesting unscreened local moments.
The bottom  panel  shows the
Korringa-Shiba ratio, defined in Eqn. (\ref{eqn:shiba}). Below $T_{coh}$
it approaches one, and above $T_{coh}$ it is larger than one and increases with $U$.
}
\label{fig:nmr}
\end{figure}

{\it Nuclear spin relaxation rate.}
This is given by
\begin{equation}
{ 1 \over T_1 T} = A^2 \lim_{\omega \to 0}  {\chi''(T,\omega) \over  \omega}
\label{eqn:1t1}
\end{equation}
and in a  Fermi liquid this quantity is independent of temperature (Korringa).  Note that there is a relationship 
to the electron spin relaxation rate $\Gamma$ defined in equation (\ref{eq:eq1}), $T_1 \sim \Gamma/(T \chi'(T,\omega=0))$. 
Hence, $T_1$ being independent of $T$, which is sometimes associated with quantum criticality \cite{Sachdev:2000,Kitagawa:2005},
is not the same  as $\Gamma$ being linear in $T$, if the temperature dependence of the dc susceptibility is
significant (as it is here).  The top panel of figure \ref{fig:nmr} shows $1/(T_1 T)$  as a function of temperature for a range of $U$ values.
Below the coherence temperature for the metallic phase it is independent of temperature, as expected. Its magnitude is 
significantly enhanced as the Mott insulator is approached.

For comparison, we note that Zitko, Osolin, and Jeglic \cite{Zitko15} calculated $\chi''(T,\omega) /  \omega$ 
for a doped  Hubbard model at filling $n=0.8$ using the numerical normalization group as an impurity solver in DMFT.
They found that $1/T_1$ was a non-monotonic function of temperature and increased by up to two orders of magnitude as $U/W$ 
increased from 0 to 4, and was weakly temperature dependent in the bad metal regime. 

{\it NMR Knight shift.}
In a lattice system this is given by $K(T) = A \chi'(\vec{q}=0,\omega=0)$, where  $\vec{q}$ is the wave vector 
and $A$ is the hyperfine coupling.  Note that the right hand side is not the same quantity as the local spin 
susceptibility, $\chi'(\omega=0) \equiv \sum_{\vec{q} }\chi'(\vec{q},\omega=0)$, that is our focus here.
Nevertheless, for reasons of simplicity, here we do not consider this difference.   The middle panel of figure \ref{fig:nmr} shows 
the static local susceptibility as a function of temperature for a range of $U$ values.  Here we work with 
units such that $A=1$.  Note that in the metallic phase as the temperature increases there is a crossover from 
a temperature-independent value at low temperatures, characteristic of a Fermi liquid, to a Curie form $1/(4T)$, 
characteristic of localized non-interacting spins. 

{\it Korringa-Shiba relation.}
In a simple Fermi liquid the following dimensionless ratio is unity in the absence of vertex corrections \cite{Yusuf:2009},
\begin{equation}
\kappa(T) \equiv 
\lim_{\omega \to 0}
{\chi''(T,\omega) \over  2 \pi \omega \chi'(T,\omega)^2}.
\label{eqn:shiba}
\end{equation}
Shiba \cite{Shiba:1975} showed that for the single Anderson impurity model $\kappa(T)=1$ in the Kondo regime. 
Values of $\kappa$ larger and less than one are often associated respectively  with antiferromagnetic and 
ferromagnetic fluctuations.  In Figure \ref{fig:nmr} we plot this ratio as a function of temperature for a 
range of $U$, and find that it can be much larger than unity and increases as the Mott insulating phase is 
approached from the metallic side.  However, in the Fermi liquid regime, $\kappa$ is close to one.

\section{Absence of quasi-particles in the quantum critical region}

\begin{figure}[htb]
\centering
\includegraphics[angle=0,width=1.0\columnwidth]{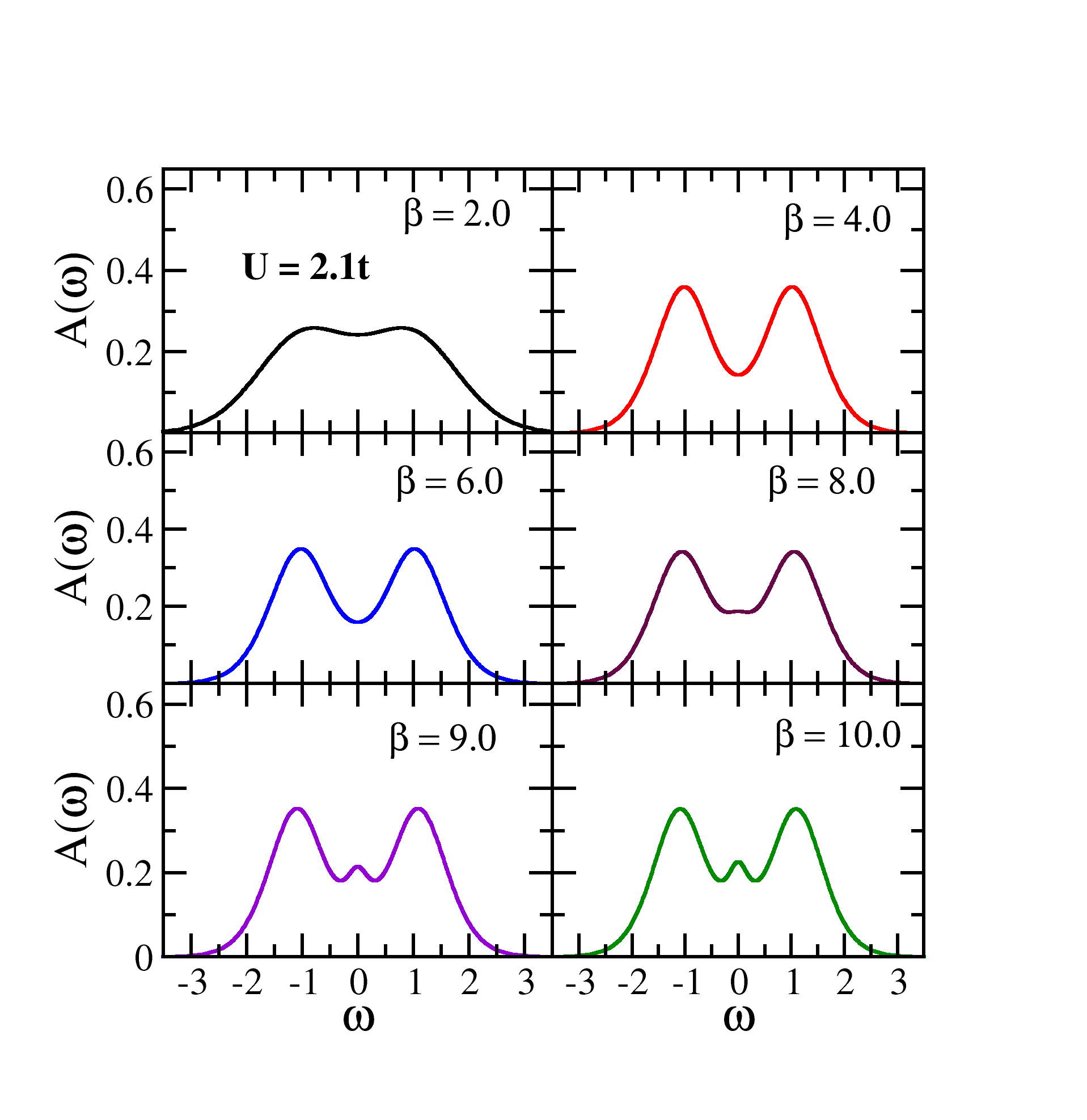}
\caption{
One electron spectral density for $U/t=2.10$ for different temperatures
in the quantum critical region. Compare Fig. 7 in Ref. \onlinecite{Vucicevic:2013}.
 Absence of a quasi-particle peak at $\omega=0$ is characteristic of a bad metal \cite{Merino:2000}.
}
\label{fig:spectral}
\end{figure}

In Figure \ref{fig:spectral} we show the one electron spectral density for $U/t=2.10$.
It can be seen that for temperatures in the quantum critical region  there is an absence of
the  quasi-particle peak at $\omega$ that is characteristic of a Fermi liquid.

\section{Relevance to experimental results for organic charge transfer salts}

The materials that are arguably closest to the model considered here are organic charge transfer salts, e.g.
$\kappa$-(BEDT-TTF)$_2$X. They can be modeled in terms of an effective Hamiltonian that is a single band 
Hubbard model on an anisotropic triangular lattice at half filling \cite{Powell:2011}.  As the pressure 
increases these materials undergo a first order phase transition from a Mott insulator to a Fermi liquid metal.
It has been found that DMFT describes the crossover from a coherent Fermi liquid to  bad metallic state with 
increasing temperature~\cite{Merino:2000}. Furthermore, DMFT  gives a quantitative description of the 
resistivity~\cite{Limelette:2003} and the frequency dependent optical conductivity~\cite{Merino:2008} for 
these organics.  Near the critical point, some signatures of  critical behavior have been reported in the 
conductivity \cite{Kagawa:2005,Furukawa:2015} and NMR \cite{Kagawa:2009}.  For a diverse set of  
$\kappa$-(BEDT-TTF)$_2$X 
above some temperature of the order $T_{b} \sim $ 50 K, the NMR relaxation
rate becomes roughly independent of temperature \cite{Powell:2009}.
Broadly, this is consistent with the quantum criticality discussed here.
On the other hand, there are alternative explanations in terms of short-range antiferromagnetic
spin fluctuations \cite{Powell:2009}, and the experiments cover a relatively narrow temperature range, roughly 
50-300 K, which is not even a single decade.  Our results compare well qualitatively with experimental
results for 
$\kappa$-(BEDT-TTF)$_{2}{\text{Ag(CN)}}_{3}$
(see Figure 3(a) in Reference \onlinecite{Shimizu:2016}).
and $\kappa$-(BEDT-TTF)$_{4}{\text{Hg}}_{2.89}{\text{Br}}_{8}$
(see Figure 3(c) in Reference \onlinecite{Eto:2010}).
(Although, it should be noted that the latter material
has been suggested to be doped away from half filling.)
As the pressure increases $1/(T_1 T)$ decreases by more than an order of magnitude.
It smoothly crosses over from a form that is monotonically decreasing with temperature
above about 10 K at low pressures to weak temperature dependence (Korringa) at higher 
pressures. 
In several organics the Korringa ratio is observed to be temperature
dependent with  large values of order ten \cite{Desoto,Itaya:2009}.
We hope our results will stimulate new experiments.

\section{Discussion and Conclusions}

The observed quantum critical scaling in the dynamical spin susceptibility above 
$T_c$ is what one expects to be associated with a zero temperature quantum critical 
point \cite{Vucicevic:2013,David}. 
This quantum critical region might be extended down to zero temperature by varying some parameter 
such as doping.  For example, the crossover scale between the region of linear resistivity and 
the low temperature Fermi liquid regime was found to vanish as half filling is 
approached\cite{m_jarrell_94,t_pruschke_95} (compare Figure 9 in Ref. \onlinecite{Vucicevic:2013}). 
Recent exact results on doped Mott insulators within DMFT\cite{David} identified a continuous quantum phase
transition from the metal to Mott insulator phase through the absence of a 
co-existence region 
in the limit of particle-hole asymmetry parameter $1-\frac{2\mu}{U} \rightarrow 1$.  This implies
that the bottom of the quantum critical fan associated with Mott quantum criticality can be pushed
down to zero temperature in
this limit. CTQMC results by  Vucicevic et al\cite{Vuvcivcevic:2015} on the doped Mott insulator, indeed, support such an implication,
since they show that the quantum critical scaling of the DC conductivity extends to much lower temperatures
than what was found in the symmetric case.

Quantum criticality means that other dynamical response functions such as the frequency-dependent 
conductivity should also exhibit $\omega/T$ scaling.  We observed that in the quantum critical 
region the one-particle Green's function exhibits $\omega/T$ scaling and hence one could expect 
the same scaling for the optical conductivity in the critical region since only the zeroth order 
bubble survives in the infinite dimensional limit.  


The nature of the quantum critical point is also of interest.  In strongly correlated electronic systems two 
types of quantum criticality are most often discussed\cite{q_si_10,q_si_14}: a local quantum quantum critical 
point associated with the destruction of Kondo screening\cite{q_si_10}, and a Moriya-Hertz-Millis (see, 
e.g., Reference \onlinecite{j_lohneysen_07}) critical point associated with the destruction of a spin density wave 
or some other ordered phase.  These two scenarios for electron models may be distinguished by the fact that 
the point where the Kondo screening vanishes at zero temperature coincides wtih the QCP in the local 
model, but not in the Moriya-Hertz-Millis scenario.  In addition, the former displays $\omega/T$ scaling in 
the quantum critical region, whereas the latter displays $\left(\omega/T \right)^{1-\theta}$ scaling with 
$\theta >0\cite{q_si_10}$.  As noted previously\cite{m_jarrell_94,t_pruschke_95}, the crossover scale 
between the region of anomalous transport, linear resistivity, and the Fermi liquid vanishes as the doping 
$x\to 0$, and the slope of the linear in temperature resistivity varies like $1/x$, both strongly suggesting 
that the QCP is at half filling (but not necessarily at zero chemical potential, since it varies discontinuously 
with the opening of the Mott gap).  These phenomena coincide with a vanishing Kondo peak in the density
of states, with width given by the crossover scale,  as $x\to 0$.  Our findings, together with these previous 
results, indicate that the QCP in this model is a local quantum critical point. 

It is interesting to speculate if the quantum criticality found here could be related to the Quantum Critical 
Point (QCP) found at finite doping in the 2D Hubbard model using Dynamical Cluster QMC 
simulations\cite{n_vidhyadhiraja_09,e_khatami_10}.  There the QCP separates the pseudogap and Fermi liquid
regions, with a large region of $\omega/T$ 
marginal Fermi liquid \cite{Varma:1989} scaling above the QCP.  Calculations are now underway to 
explore this possibility.

In summary, the main significance of this work is that it gives a concrete example of a fermion model 
which has a metallic state in proximity to a Mott insulating state and has dynamical local spin 
fluctuations that exhibit the $\omega/T$ scaling that is characteristic of local quantum 
criticality.

\begin{acknowledgments}
ND thanks CSIR and DST (India) for research funding. RHM was supported in part by an Australian 
Research Council Discovery Project. He also benefited from discussions at the Aspen Center for Physics, 
which is supported by National Science Foundation grant PHY-1066293.  MJ was funded by the NSF EPSCoR
La-SiGMA project, No. EPS-1003897.
We thank Vladimir Dobrosavljevic, Steve Kivelson, Jure  Kokalj, H.R.\ Krishnamurthy, Alejandro Mezio, Juana Moreno, Ben Powell,
Peter Prelovsek, Sri Raghu, Qimao Si, Darko Tanaskovic, Arghya Taraphder, and Rok Zitko for helpful discussions.
Our simulations used an open source implementation~\cite{Hafer} of the hybridization expansion
continuous-time quantum Monte Carlo algorithm~\cite{Comanac} and the ALPS~\cite{Bauer}
libraries. The computational resources are provided by
the Louisiana Optical Network Initiative (LONI) and HPC@LSU.
\end{acknowledgments}

\appendix*

\section{Determination of the coherence temperature}

\begin{figure}[htb]
\begin{center}
\includegraphics[angle=0,width=0.9\columnwidth]{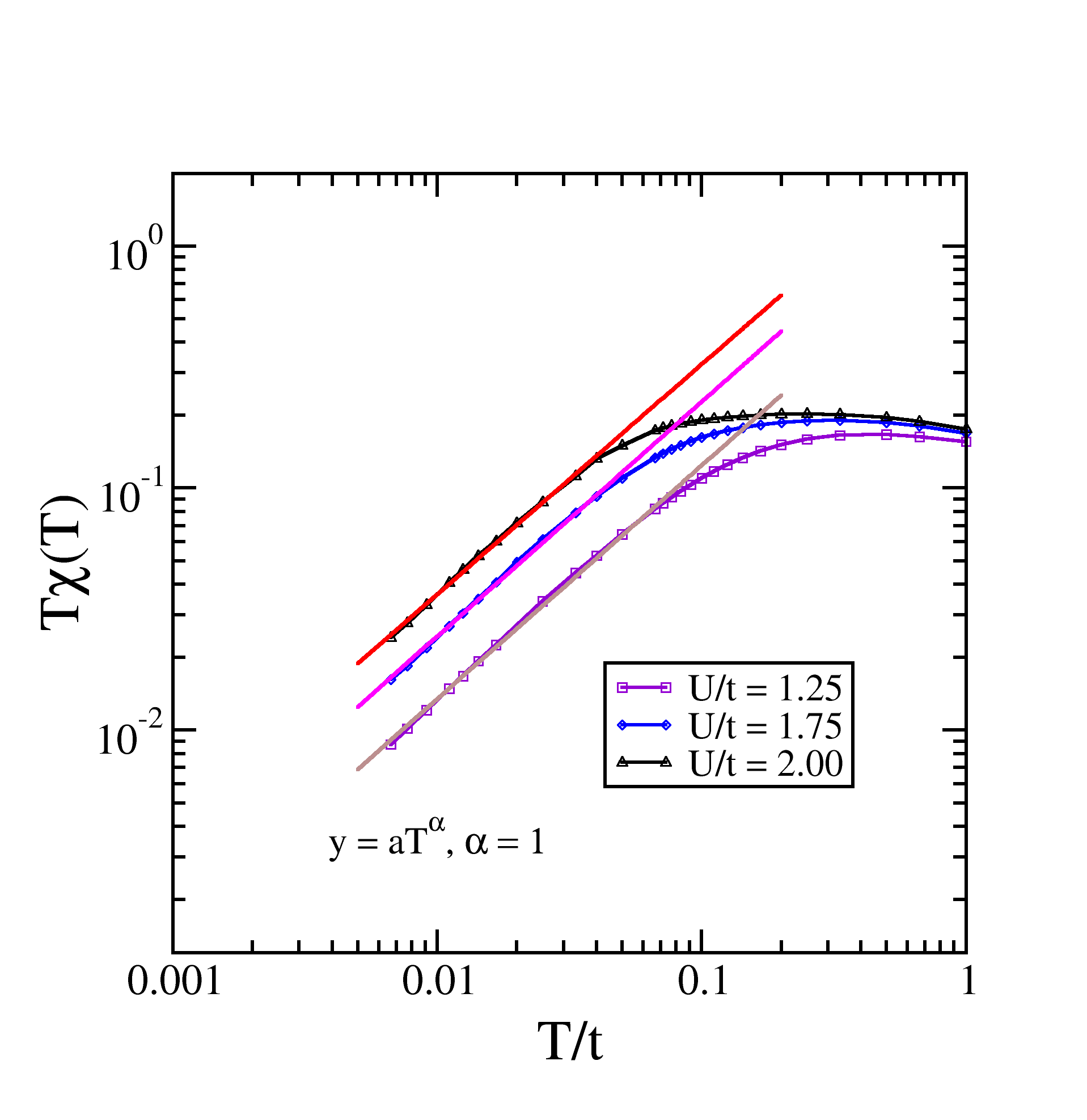}
\includegraphics[angle=0,width=0.9\columnwidth]{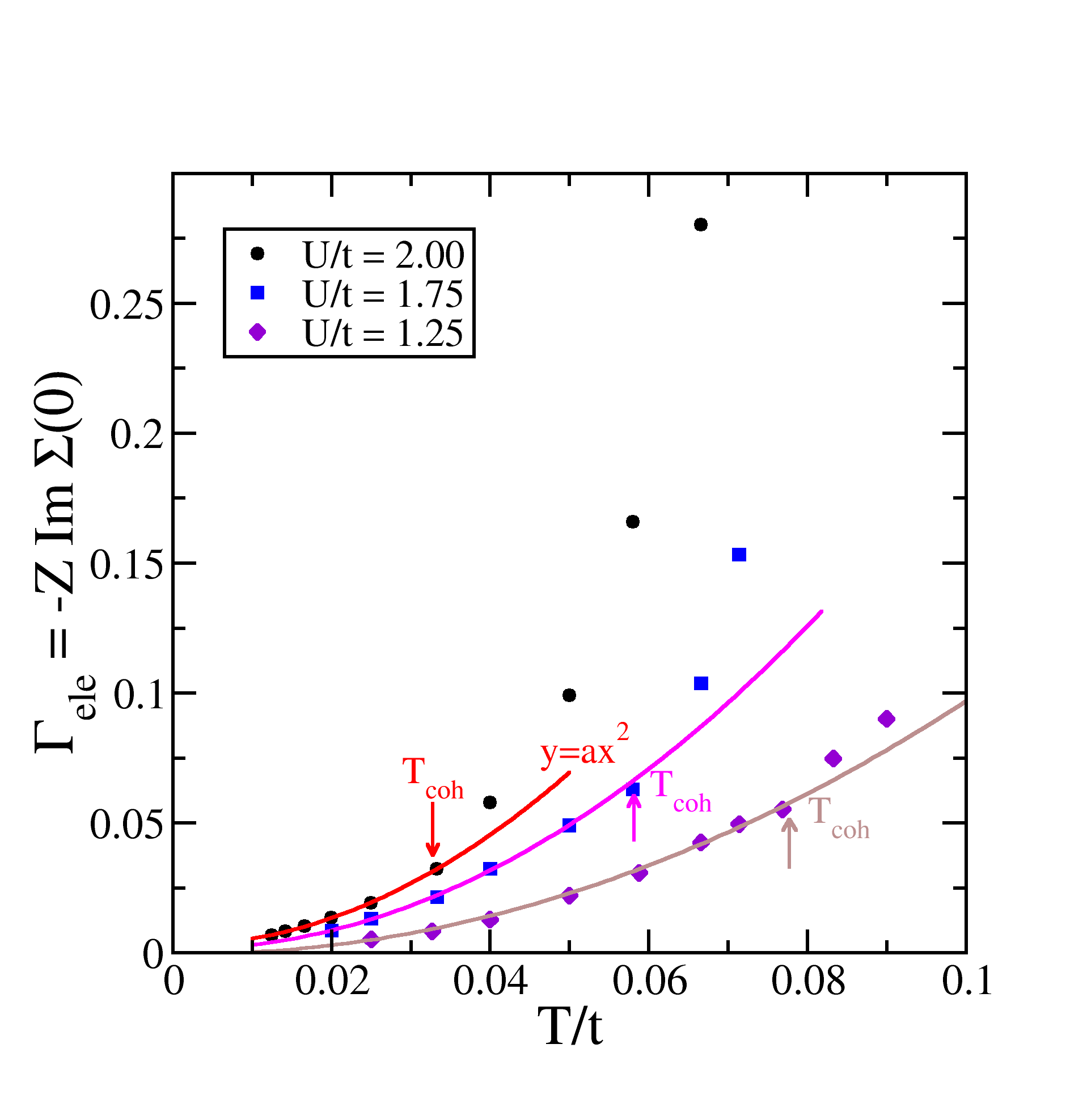}
\end{center}
\caption{
Determination of the Fermi liquid coherence temperature. Upper panel: from the static local spin susceptibility,
$T_{coh}$ is defined as the temperature at which $\chi(T)$ deviates from the temperature
independence characteristic of a Fermi liquid.
Lower panel: from the self energy for the one-electron Green's,
$T_{coh}$ is defined as the temperature at which $\Sigma''(\omega=0,T)$ deviates from the quadratic temperature
dependence characteristic of a Fermi liquid.
}
\label{fig:Tcoh}
\end{figure}

Figure \ref{fig:Tcoh} shows the Fermi liquid coherence temperature determined by
two distinct methods.
This is not necessarily the same for different properties.
Sometimes, it is smaller for  two-particle properties than for single-particle ones\cite{Mravlje}. 
For example,  in the Kondo problem the Kondo resonance exists up to temperatures of $2T_K$, while the spin susceptibility saturates to a Pauli form only below $T=0.2T_K$ .
For example, in Ref. \onlinecite{Costi} compare Figures 2, 7, and 16, which show the specific heat, spectral density, and thermopower, respectively.

\bibliographystyle{apsrev4-1}
\bibliography{badspin}

\end{document}